\documentclass[aps, prl, reprint, twocolumn, floatfix, superscriptaddress, longbibliography]{revtex4-1} 
\usepackage{graphicx}
\usepackage{epstopdf}
\usepackage{bm, amsmath, amssymb, bbm}
\usepackage{bbold}
\usepackage{float}
\usepackage{placeins}
\usepackage[dvipsnames]{xcolor}
\usepackage{siunitx}
\usepackage[T1]{fontenc}
\usepackage{enumitem} 
\usepackage{comment}
\usepackage{soul}

\begin{document}

    \title{Fully Directional Quantum-limited Phase-Preserving Amplifier}
    \author{G. Liu$^\ddagger$}
    \email{gangqiang.liu@yale.edu}
    \affiliation{Department of Applied Physics, Yale University, New Haven, CT 06520, USA}    
    \author{A. Lingenfelter$^\ddagger$}
    \affiliation{Department of Applied Physics, Yale University, New Haven, CT 06520, USA} 
    \affiliation{Department of Physics, University of Chicago, Chicago, IL 60637, USA}  
    \author{V. R. Joshi}
    \author{N. E. Frattini}
    \author{V. V. Sivak}
    \author{S. Shankar}
    \author{M. H. Devoret}
    \email{michel.devoret@yale.edu}
    \thanks{$^\ddagger$ these authors contributed equally}
    \affiliation{Department of Applied Physics, Yale University, New Haven, CT 06520, USA}
    \date{\today}
    
\begin{abstract}

We present a way to achieve fully directional, quantum-limited phase-preserving amplification in a four-port, four-mode superconducting Josephson circuit by utilizing interference between six parametric processes that couple all four modes. Full directionality, defined as the reverse isolation surpassing forward gain between the matched input and output ports of the amplifier, ensures its robustness against impedance mismatch that might be present at its output port during applications. Unlike existing directional phase-preserving amplifiers, both the minimal back-action and the quantum-limited added noise of this amplifier remains unaffected by noise incident on its output port. In addition, the matched input and output ports allow direct on-chip integration of these amplifiers with other circuit QED components, facilitating scaling up of superconducting quantum processors. 

\end{abstract}
\maketitle

\subsection{Introduction} 

Directional, quantum-limited signal amplification belongs to the cadre of basic quantum information processing tasks. In superconducting circuits, it is commonly performed by a combination of ferrite-based circulators and reflection Josephson parametric amplifiers \cite{Johnson2011, Riste2012, Jeffrey2014, Walter2017}. Even though these amplifiers work very close to the quantum limit, photon loss in circulators and other associated components in the signal pathway significantly reduce the quantum efficiency of this amplification process. Furthermore, the strong magnetic field needed for the operation of ferrite-based circulators and their bulkiness make them difficult to integrate with superconducting circuits. Therefore, a question naturally rises: is it possible to achieve directional quantum-limited amplification without using ferrite-based circulators?

In recent theoretical works by Ranzani et al. \cite{Ranzani2014,Ranzani2015}, it is shown that in a parametrically coupled system, nonreciprocity can be generated from the dissipation of ancillary modes and the interference of multiple coupling paths that connect the input mode to the output mode. Following this concept, both phase-preserving and phase-sensitive directional quantum-limited Josephson parametric amplifier have been demonstrated \cite{Sliwa2015, Lecocq2017, Abdo2018, Lecocq2020}. Meanwhile, Metelmann et al. have shown that any coherent coupling can be made directional by balancing it with a dissipative process \cite{Metelmann2015, Metelmann2017}. This method has been employed in the demonstration of nonreciprocity in both optical domain and microwave domain with optomechanics \cite{Ruesink2016,  Peterson2017, Fang2017, Bernier2017, Barzanjeh2017, Ruesink2018, Malz2018, deLepinay2019, Herrmann2022}.

However, in these directional amplifiers, there is either unity transmission in the reverse direction or amplification in reflection from the output port \cite{Abdo2014, Sliwa2015, Lecocq2017, Metelmann2017, Abdo2018, Malz2018, Lecocq2020}. To ensure that only vacuum noise goes back to the signal source (i.e. minimal back-action) and that only quantum-limited noise is added to the output signal, it requires that only vacuum noise can enter the output port of the amplifier. Unfortunately, this is hardly the case in experiments; the amplifier output port is typically connected to parts of the apparatus which are thermalized at temperatures much higher than the vacuum noise effective temperature of the frequency band of interest. Thermal photons are then emitted towards the directional amplifier, leading to unwanted back-action and added noise. Therefore, it is desirable to build directional amplifier with matched input and output ports, as well as sufficient reverse isolation between them. Most importantly, the reverse isolation needs to exceed the forward gain. This ensures robustness of its minimal back-action and quantum-limited noise performances against any impedance mismatch and noise that might be present on its output port during applications. We call such amplifiers the {\it{fully}} directional quantum-limited amplifiers.    

In this work, we show that fully directional quantum-limited phase-preserving amplifiers can be built with 4-port 4-mode systems that have properly arranged two-mode squeezing and frequency conversion couplings between the modes. To reach this result, we first derive the minimum scattering matrices of fully directional quantum-limited phase-preserving amplifiers. We find that there exists two such scattering matrices which represent 4-port systems. From these scattering matrices, we then generate the coupling matrices between the modes. These coupling matrices in turn provide the full implementation guideline for these amplifiers. We also theoretically investigate the performance of these amplifiers under practical operating conditions and find they are robust against imperfection in parametric couplings and input signal detuning.

\section{Minimal scattering matrix of a fully-directional quantum-limited amplifier}

We start with the scattering matrix representation of a linear amplifier. For a linear amplifier with $N$ ports, the input and output signals are related by 
\begin{align}
    \label{eq:S_general}
    \begin{bmatrix}
    a^{\rm{out}}_1 \\
    a^{\rm{out \dagger}}_1 \\
    \vdots \\
    a^{\rm{out}}_N \\
    a^{\rm{out \dagger}}_N \\    
    \end{bmatrix} = \tilde{S} 
    \begin{bmatrix}
    a^{\rm{in}}_1 \\
    a^{\rm{in \dagger}}_1 \\
    \vdots \\
    a^{\rm{in}}_N \\
    a^{\rm{in \dagger}}_N \\    
    \end{bmatrix}
\end{align}
where $a^{\rm{in}}_n$ ($a^{\rm{in} \dagger}_n$) and $a^{\rm{out}}_n$ ($a^{\rm{out} \dagger}_n$) are field operators of the input and output signals on the $n$-th port of the amplifier, and $\tilde{S}$ is the $2N$-by-$2N$ scattering matrix. The reason for separating $a$ and $a^\dagger$ for each port is that the amplifier might not a priori amplify the two quadratures by the same coefficient.

The requirement for performing fully directional quantum-limited phase-preserving amplification with such an amplifier translates into the following requirements on its scattering matrix, 
\begin{enumerate}[label=(\roman*)]
    \item {The scattering matrix can always be block-diagonalized on a proper mode basis,
    \begin{align}
    \label{eq:S_pp}
    L\tilde{S}L^{-1} =  
    \begin{bmatrix}
    S & 0 \\
    0 & S^*
    \end{bmatrix}
    \end{align}}
    where $L$ is the linear transformation from the mode basis in Eq.~(\ref{eq:S_general}) to the proper one. $S$ is a $N$-by-$N$ matrix which is often referred to as the scattering matrix even though it might mix $a_n$ and $a^\dagger_m$. The property of these two sub-matrices being conjugate ensures phase-preserving amplification. In the graph-based notation of the scattering matrix \cite{Ranzani2015,Naaman2022}, this property ensures that the full graph can be decomposed into two sub-graphs.
    \item The scattering matrix is symplectic,
    \begin{align}
    \label{eq:S_symp}
        \tilde{S}^{T} \cdot J \cdot \tilde{S} = J
    \end{align}
    where $J$ is the symplectic matrix
    \begin{align}
    J=
    \begin{bmatrix}
        0 & 1 & & & \\
        -1 & 0 & & & \\
        & & \ddots & & \\
        & & & 0 & 1 \\
        & & & -1 & 0
    \end{bmatrix}.
    \end{align}
   The meaning of Eq.~(\ref{eq:S_symp}) is that the device preserves the information content of the input signal, or in other words it preserves the commutation relation of the input field operators \cite{Caves1982}.
   
    \item The sub-scattering matrix between the input port (Port 1) and output port (Port 2) should take the form
    \begin{align}
    \label{eq:S_directional}
        [S]_{1,2} = \begin{bmatrix}
        0 & 0 \\
        \sqrt{G}e^{i\phi} & 0
        \end{bmatrix}
    \end{align}
    where $S_{21} = \sqrt{G}e^{i\phi}$ is the gain from the input to the output port. This stipulates that both the input and output ports are matched, and the reverse isolation ($1/|S_{12}|$) is infinity.
    
    \item Scattering matrix elements that couple other ports to the input and output ports should satisfy the following condition, 
    \begin{align}
    \label{eq:S_QL}
        \sum_{p=3}^{N} |S_{1p}|^2 = 1, \quad \sum_{p=3}^{N} |S_{2p}|^2 \approx G \ \rm{for} \ G\gg 1, 
    \end{align}
    such that both the back action and added noise of the amplifier are quantum-limited when only vacuum noise enters the output port and all other auxiliary ports.
\end{enumerate}
With this set of conditions, we obtain equations for the $N^2$ complex scattering matrix elements. In solving these equations (Supplementary Material Section A), we find that solutions only exist for $N \geq 4$. Therefore, at least 4 ports -- input port, output port and two ancillary ports -- are needed to perform fully directional quantum-limited phase-preserving amplification.

\begin{figure}[htbp]
    \centering
    \includegraphics[width=1.0\linewidth]{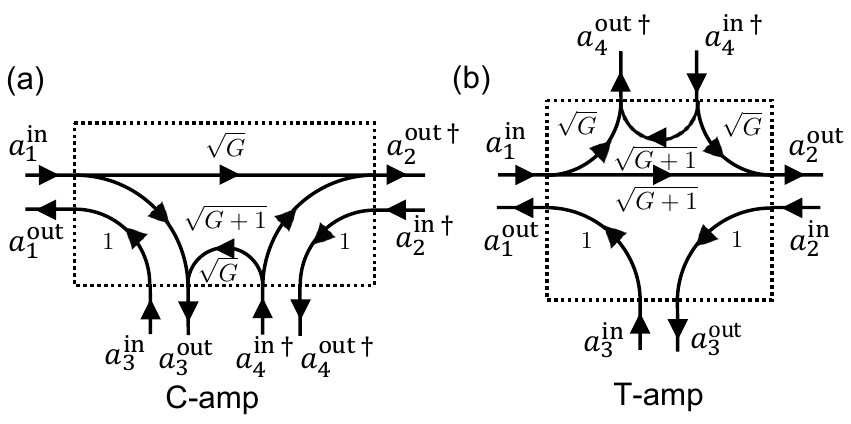}
    \caption{Scattering graphs of the two 4-port fully directional amplifiers nicknamed C-amp and T-amp. (a) C-amp has a "cis" configuration of the ancillary ports meaning they are on the same side of the wave trajectory from Port 1 to Port 2, and (b) T-amp has a "trans" configuration of the ancillary ports meaning they are on the opposite side of the wave trajectory from Port 1 to Port 2. Scattering matrix elements between two ports are represented by lines with arrow indicating wave trajectory direction between ports. The value of the scattering matrix elements is shown next to the line.}
    \label{fig:4PFDA_scattering_graph}
\end{figure}

For a 4-port system, we find that there exist two types of such 4-by-4 scattering matrices. We name the corresponding 4-port fully directional amplifiers (4PFDAs) as Cis- and Trans-amplifier (C-amp and T-amp) based on the 'topology' of their scattering graph (see Fig.~\ref{fig:4PFDA_scattering_graph}), respectively. Minimal form of these scattering matrices are, for the C-amp
\begin{align}
\label{eq:S_C-amp_QL_min}
    S_{C} = {\begin{bmatrix}
    0 & 0 & 1 & 0 \\
    \sqrt{G} & 0 & 0 & \sqrt{G+1} \\
    \sqrt{G+1} & 0 & 0 & \sqrt{G} \\
    0 & 1 & 0 & 0 \\
    \end{bmatrix}}
\end{align}
and for the T-amp
\begin{align}
\label{eq:S_T-amp_QL_min}
    S_{T} = {\begin{bmatrix}
    0 & 0 & 1 & 0 \\
    \sqrt{G+1} & 0 & 0 & \sqrt{G} \\
    0 & 1 & 0 & 0 \\
    \sqrt{G} & 0 & 0 & \sqrt{G+1} \\
    \end{bmatrix}}.
\end{align}
where $G \ge 0$. The corresponding mode bases are
\begin{align}
    A_{C}^{\rm{in,out}} = \begin{bmatrix}
        a_1 \\
        a^{\dagger}_2 \\
        a_3 \\
        a^{\dagger}_4
        \end{bmatrix}^{\rm{in,out}}, \quad
    A_{T}^{\rm{in,out}} = \begin{bmatrix}
        a_1 \\
        a_2 \\
        a_3 \\
        a^{\dagger}_4
        \end{bmatrix}^{\rm{in,out}}
\end{align}
respectively. Note the matrices have been made positive and real after a change of phase of the incoming and outgoing waves. In general, scattering matrices depend on input signal frequency while Eq.~(\ref{eq:S_C-amp_QL_min}) and (\ref{eq:S_T-amp_QL_min}) represent the special case of resonant input signal. These two scattering matrices are graphically represented in Fig.~\ref{fig:4PFDA_scattering_graph}.

There are important differences between these two types of amplifiers, although both provide same scattering relation between the input and output ports in the high gain limit ($G \gg 1$). First, all ports of C-amp are matched while T-amp has one ancillary port with reflection gain. Second, from the mode basis, one notices that the conjugate frequency components of the input and output are coupled in C-amp while the same frequency components are coupled in T-amp. The two auxiliary ports play crucial roles: when they are terminated with matched cold load, they serve as 'dumps' for signal entering the device from the output port, therefore providing isolation from output port to input port. Meanwhile, their conjugation also provides vacuum noise going to the input port as required for minimum quantum back-action.  

\section{Mode coupling of the 4PFDA}

With the minimal scattering matrices determined, we now discuss how to obtain these scattering matrices from a coupled-mode system. In general, there can be more modes than ports in such systems, which becomes necessary when engineering to increase the bandwidth of the amplifier \cite{Naaman2022}. We focus here on the case of minimal number of required modes. We therefore consider a $N$-port $N$-mode system.  The scattering matrix of such a system is given by its coupling to the environment and between the modes as (see derivation in SM) 
\begin{equation}
    \label{eq:Z_to_S}
    S = (\Sigma + M)^{-1}(\Sigma - M)
\end{equation}
with 
\begin{equation}
    \label{eq:Zc}
    \Sigma =
    \begin{bmatrix}
    \kappa_1/2 & & & \\
    & \cdot & & \\
    &  & \cdot & \\
    & & & \kappa_N/2
    \end{bmatrix}
\end{equation}
which represents the amplitude damping rate of the modes due to their coupling to the environment, and
\begin{equation}
    \label{eq:Z}
    M =
    \begin{bmatrix}
    i\Delta_1 & M_{12} & \cdots & M_{1N} \\
    M_{21} & i\Delta_2 & \cdots & M_{2N} \\
    \vdots &  \vdots & \cdots & \vdots \\
    M_{N1} & M_{N2} & \cdots & i\Delta_N
    \end{bmatrix}
\end{equation}
which represents the coupling between the modes ($M_{mn}$) and to the external drives ($i\Delta_n$) with $\Delta_n=\omega^{\rm{in}}_n - \omega_n$. Eq.~(\ref{eq:Z_to_S}) can be viewed as the generalized reflection coefficient between two systems of different impedances. Given the scattering matrix $S$, one can obtain the coupling between the modes from Eq.~(\ref{eq:Z_to_S}) as
\begin{equation}
    \label{eq:S_to_Z}
    M = \Sigma(I-S)(I+S)^{-1}
\end{equation}
where $I$ is the identity matrix.

From the scattering matrices given in Eq.~(\ref{eq:S_C-amp_QL_min}) and (\ref{eq:S_T-amp_QL_min}), we obtain the coupling Hamiltonians of these two amplifiers as
\begin{widetext}
\begin{align}
\label{eq:C_hamiltonian}
    H_C/\hbar = & \frac{1}{2} \left( g_{12} a_1 a_2 + g_{13} a_1 a^\dagger_3 + g_{14} a_1 a_4 + g_{23} a_2 a_3 + g_{24} a_2 a^\dagger_4 + g_{34} a_3 a_4 \right) + h.c.
\end{align}
\begin{align}
\label{eq:T_hamiltonian}
    H_T/\hbar = & \frac{1}{2} \left( g_{12} a_1 a^\dagger_2 + g_{13} a_1 a^\dagger_3 + g_{14} a_1 a_4 + g_{23} a_2 a^\dagger_3 + g_{24} a_2 a_4 + g_{34} a_3 a^\dagger_4 \right) + h.c.
\end{align}
\end{widetext}
where, 
\begin{align}
    g_{mn} = \pm i \sqrt{\kappa_m \kappa_n} \frac{\sqrt{G-1}}{\sqrt{G}+1}
\end{align}
for terms with $a_m a_n$, and 
\begin{align}
    g_{mn} = \pm i \sqrt{\kappa_m \kappa_n}
\end{align}
for terms with $a_m a^\dagger_n$, which correspond to {\it{two-mode squeezing}} and {\it{frequency conversion}} between the modes, respectively. The sign of each term, i.e. the phase of each coupling, is uniquely determined by the scattering matrix. The two-mode squeezing coupling alone leads to quantum-limited phase-preserving amplification of input signals to the two modes with photon number gain of $G$ in reflection ($G-1$ in transmission). The frequency conversion coupling leads to {\it{perfect}} photon conversion, henceforth denoted as $C=1$, between the two modes. 

\begin{figure}[htbp]
    \centering
    \includegraphics[width=1.0\linewidth]{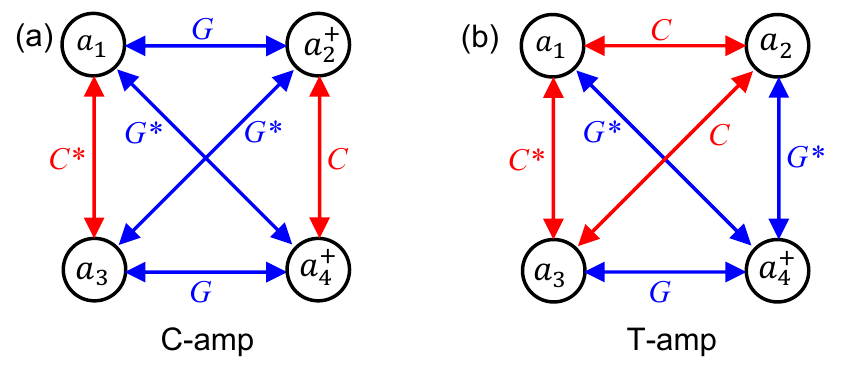}
    \caption{All-parametric implementation of the two 4-port fully directional amplifiers in Fig.~\ref{fig:4PFDA_scattering_graph}. (a) Parametric coupling graph of C-amp. There are two conversion couplings ($C$ or $C^*$, red) and four gain couplings ($G$ or $G^*$, blue). (b) Parametric coupling graph of T-amp. There are three conversion couplings and three gain couplings. The phase of each individual coupling is either $\pi/2$ for $C$ and $G$, or $-\pi/2$ for $C^*$ and $G^*$. }
    \label{fig:4PFDA_coupling graph}
\end{figure}

These coupling Hamiltonians are graphically illustrated in Fig.~\ref{fig:4PFDA_coupling graph}. First, one notices that the total phases of the three couplings associated with each mode is either $\pi/2$ or $-\pi/2$. Furthermore, each closed loop of three coupled modes forms a 3-port directional amplifier or circulator which have been demonstrated in previous works \cite{Sliwa2015, Lecocq2017, Barzanjeh2017, Fang2017, Herrmann2022}. In fact, the coupling graphs of the C-amp and T-amp are equivalent to the two different ways of connecting a 3-port directional amplifier to a 3-port circulator. See Supplemental Material for more detailed discussions. 

In Ref.~\cite{Metelmann2017}, a 4-port 4-mode system with 5 couplings was proposed for quantum-limited directional phase-preserving amplification. However, the input port of such a system would produce unity reflection, which is undesired. Such reflection could lead to unwanted back-action on the signal source. Therefore, a 4-port 4-mode system with 6 properly arranged couplings represents the minimum construction of a fully directional quantum-limited phase-preserving amplifier. On the other hand, as shown in Ref.~\cite{Chien2020}, a fully directional phase-sensitive amplifier would require a 3-port 3-mode system with 5 couplings.

\section{Effects of imperfect conversion in 4PFDA}
So far, we have only discussed the performance of these amplifiers under ideal situation: perfect frequency conversion and resonant input signal. In this section, we will show that even under practical operating conditions -- imperfect frequency conversion and detuned input signal -- these amplifiers give superior performance compared to existing directional amplifiers based on interference between multiple parametric processes. 

\begin{figure}[htbp]
    \centering
    \includegraphics[width=1.0\linewidth]{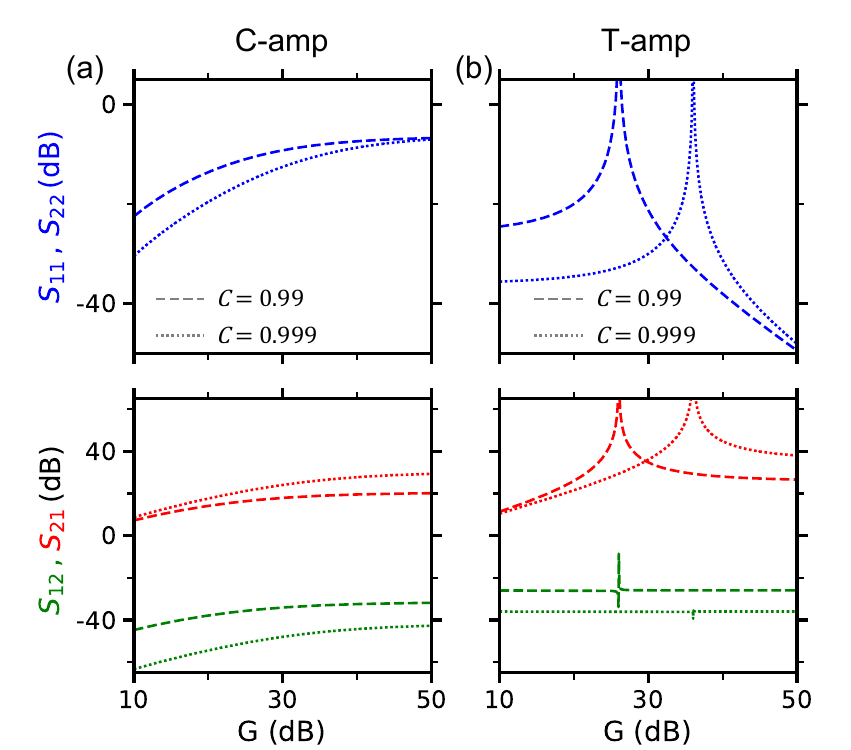}
    \caption{Scattering matrix elements of C-amp and T-amp for imperfect conversion. Input and output matching ($S_{11}$ and $S_{22}$, blue), forward gain ($S_{21}$, red) and reverse isolation ($S_{12}$, green) for C-amp in panel (a), and for T-amp in panel (b) with respect to gain of single coupling ($G$) for $C=0.99$ (dashed) and $C=0.999$ (dotted).}
    \label{fig:S_vs_G} 
\end{figure}

First of all, we investigate the effect of imperfection in frequency conversion between the modes on the performance of the amplifiers. In general, with imperfect conversion there will be reflection on the input and output ports. In Fig.~\ref{fig:S_vs_G}, we show the scattering matrix elements between the input and output ports of the C-amp and T-amp versus two-mode squeezing gain for frequency conversion of $C=0.99$ and $C=0.999$ between each pair of modes, which are readily achievable in experiments. 

For C-amp (Fig.~\ref{fig:S_vs_G}(a)), both reflection ($S_{11}$, $S_{22}$) and transmission in reverse direction ($S_{12}$) increases as the gain coupling between paired modes increases, and eventually saturates in high gain limit ($G>40$~dB). As expected, better amplification in forward direction ($S_{21}$) and isolation in reverse direction ($S_{12}$) are achieved with better conversion between the modes. With readily achievable conversion of $C=0.99$ between modes, this amplifier is matched at the -15 dB level on both input and output port while providing 20 dB forward gain. To achieve this level of input matching and gain with a single-parametric reflection amplifier, at least two circulators are needed in front of the amplifier.  

For T-amp (Fig.~\ref{fig:S_vs_G}(b)), we observe very different behaviors. Both the reflections and forward gain start to increase rapidly as gain coupling between the paired modes increases above $4/(1-C)$. However, impedance matching of both the input and output ports remain better than -20~dB while the forward gain approaches 20~dB. This performance can be readily achieved with two-mode squeezing gain of 17 dB for both frequency conversion values. Furthermore, the isolation in reverse direction is insensitive to the two-mode squeezing gain strength.

\section{Frequency dependence of 4PFDA performance}
\begin{figure}[htbp]
    \centering
    \includegraphics[width=1.0\linewidth]{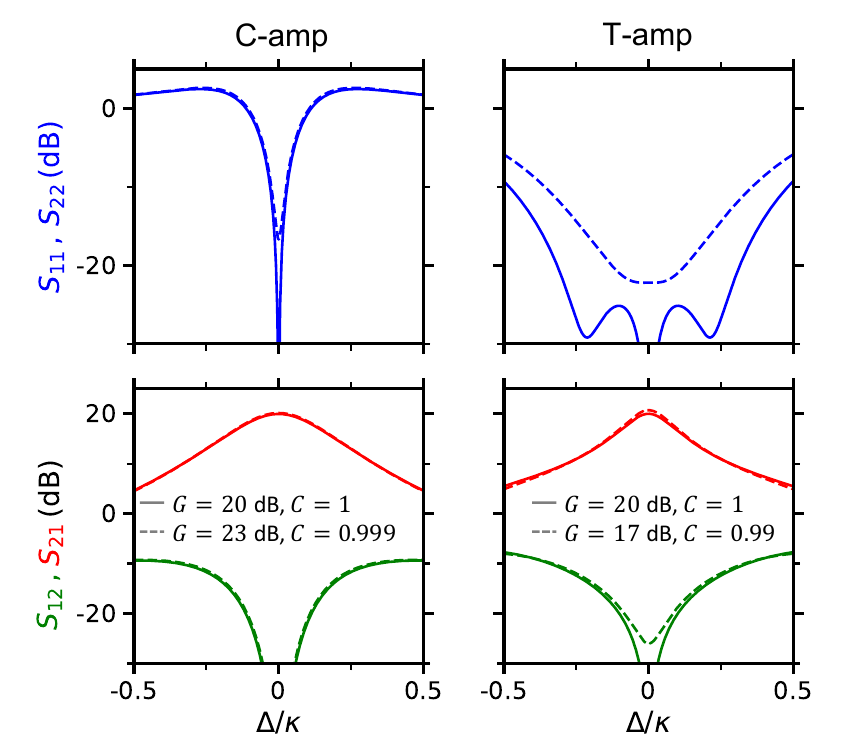}
    \caption{Frequency dependence of scattering matrix elements of C-amp and T-amp. Input and output matching ($S_{11}$ and $S_{22}$, blue), forward gain ($S_{21}$, red) and reverse isolation ($S_{12}$, green) of C-amp in panel (a), and T-amp in panel (b) with respect to reduced detuning $\Delta/\kappa$. Solid lines represent results for the case with gain couplings of $G=20$~dB and conversion couplings of $C=1$ between the modes. Dashed lines in (a) represent results for gain couplings of $G=23$~dB and conversion couplings of $C=0.999$. Dashed lines in (b) represent results for gain couplings of $G=20$~dB and conversion couplings of $C=0.99$. Note that in C-amp, much higher two-mode squeezing gain would be needed for 20 dB forward gain if C = 0.99 instead of 0.999.}
    \label{fig:4PFDA-realistic} 
\end{figure}

Now we show the amplifier performance for detuned input signal. The scattering matrix for detuned input signal is derived from the Hamiltonian of the amplifiers together with Eq.~(\ref{eq:Z_to_S}). As shown in Supplemental Material, they are complex functions of the signal detuning. In Fig.~\ref{fig:4PFDA-realistic}, we show the four scattering matrix elements between the input and output ports for frequency conversion strengths $C=1$ and $C=0.99$, while keeping the forward gain $|S_{21}|^2 = 20$~dB for resonant input signal. Without losing generality, we assume all modes have the same linewidth $\kappa$.

The most striking difference between these two amplifiers is the frequency dependence of their matching conditions (top row of Fig.~\ref{fig:4PFDA-realistic}). For C-amp, impedance matching rapidly degrades with signal detuning. For T-amp, good impedance matching can be achieved over a much larger range of detuning. The robustness of impedance matching over signal detuning in T-amp comes from the fact that two of three parametric processes that input and output modes directly participate are frequency conversions, which diminish reflections. In contrast, in C-amp two of these three parametric processes are amplifications which generate amplified signal in reflection. 

Forward gain and reverse isolation of these two amplifiers have similar frequency dependence (bottom row of Fig.~\ref{fig:4PFDA-realistic}), except that better reverse isolation is achieved in C-amp for near resonance signal. Overall, T-amp appears to be easier to operate as it requires less gain and conversion to achieve the same forward gain as C-amp.

\section{Bandwidth of 4PFDA}
A useful metric for characterizing the frequency dependent performance of an amplifier is its bandwidth, which is typically defined as the frequency range over which its gain drop by 3 dB from the desired value. For a resonator-based single-pump amplifier, such as the 2-port, 2-mode phase-preserving amplifier, its bandwidth ($B$) is related to the geometric mean of the linewidth of its two modes ($\Bar{\kappa}$) and gain by the relation $\Bar{\kappa} \approx B \sqrt{G}$, which is known as the fixed gain-bandwidth product \cite{Bergeal2010}.

For amplifiers involving multiple parametric processes such as the C-amp and T-amp, because multiple scattering parameters are relevant to their applications, their bandwidth should be defined as the frequency range over which {\it{all}} these scattering parameters maintain desired performance. In these two amplifiers, all four scattering parameters between the input and output ports are relevant. Therefore, we define the bandwidth of the them as the frequency range over which {\textit{all the following conditions are satisfied}}: $|S_{11}|^2, |S_{22}|^2 \leq 0.01$, $|S_{12}|^2 \leq 1/G$, $|S_{21}|^2 \geq G/2$, where $G$ is the forward gain for resonant input signal. 

In general, less stringent impedance matching requirement can be used, which would result in larger bandwidth for $|S_{11}|^2$ ($|S_{22}|^2$). The requirement on reverse isolation, $|S_{12}|^2 \leq 1/G$, is to ensure that the amplifier remains stable even in the extreme case that the output signal is fully reflected back into the amplifier by defects in the following components. The bandwidth of the forward gain is defined as the standard 3-dB bandwidth.

\begin{figure}
    \centering
    \includegraphics[width=1.0\linewidth]{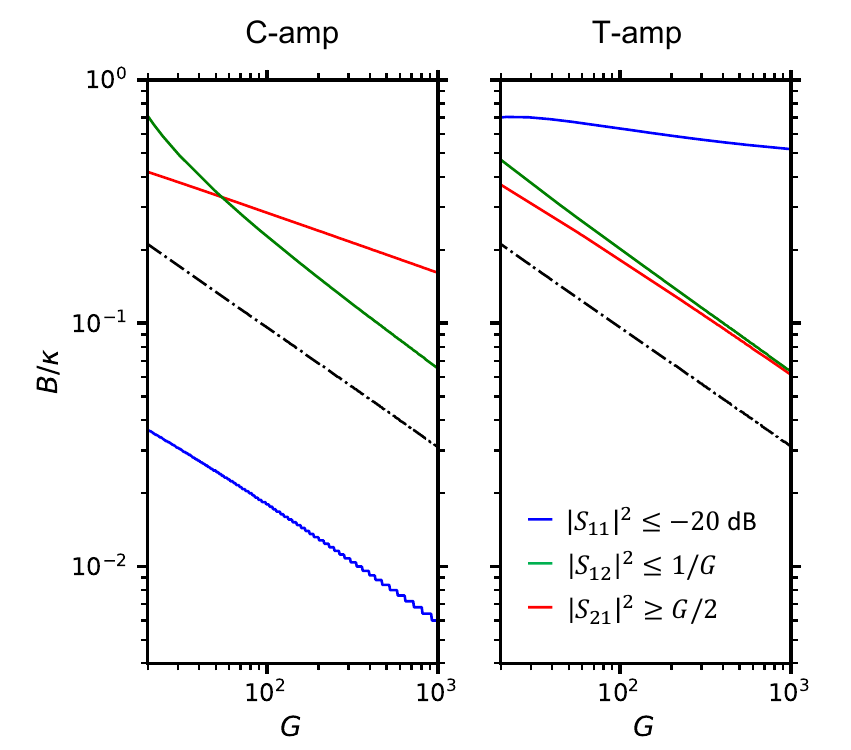}
    \caption{Bandwidth of C-amp and T-amp. Bandwidth ($B$) of scattering matrix elements for (a) C-amp and (b) T-amp as a function of gain coupling strength $G$ in the case of perfect conversion couplings ($C=1$) between modes. The bandwidth of each scattering matrix elements is defined as the frequency range over which the corresponding condition is satisfied. The dot-dashed line represent the bandwidth of forward gain ($|S_{21}|^2 \leq G/2$) of a 2-port phase-preserving amplifier.}
    \label{fig:BW} 
\end{figure}

In Fig.~\ref{fig:BW}, we show the bandwidth defined by each of these conditions versus forward gain. For C-amp, the bandwidth determined by impedance matching condition, which is nearly 10 times narrower than those determined by forward gain and reverse isolation, defines the amplifier bandwidth. For T-amp, the 3-dB bandwidth of forward gain defines the amplifier bandwidth. For both amplifiers, their 3-dB bandwidths of forward gain are approximately twice of the 3-dB bandwidth of a resonator-based single-parametric amplifier such as JPAs (dash-dotted line), while their scaling gain are similar. 

\section{Effects of imperfect impedance matching on auxiliary ports}
Impedance matching on the auxiliary ports determines reverse isolation and added noise of these amplifiers. In the discussion above, we have only considered the case in which these auxiliary parts are terminated by perfectly matched cold load. In practice, however, there will be a finite amount of impedance mismatch between the ports and its load. As long as the resultant reflection coefficient is less than $1/\sqrt{G}$, the amplifiers will remain stable. 

\section{Conclusion}
In summary, we have presented two fully directional quantum-limited phase-preserving parametric amplifiers, which have matched input and output ports and perfect reverse isolation. These amplifiers can be implemented in 4-port, 4-mode systems with properly arranged two-mode squeezing and frequency conversion couplings between the modes. The T-amp, which has frequency conversion coupling between the its input and all the other modes, gives superior operating bandwidth compared to both single-pump parametric amplifiers and 3-port directional amplifiers based on multi-parametric couplings \cite{Ranzani2015,Sliwa2015,Lecocq2017}. The C-amp, with all its ports being matched, is robust against impedance mismatch on its auxiliary ports in practical operating conditions. Although these coupling schemes suggest that 6 couplings are needed to implement such amplifiers, the number of active parametric couplings can be reduced by employing frequency degeneracy among the 4 modes such that passive frequency conversion couplings between these modes always exists. This method has been used in the implementation of directional devices in both optomechanical systems \cite{Fang2017, Herrmann2022} and superconducting circuits \cite{Abdo2014, Abdo2018, kwende2023}. These 4-port, 4-mode amplifiers are suitable for direct integration with superconducting circuits, and high efficiency readout of large scale superconducting qubits.

\section{Acknowledgement}
We acknowledge helpful discussions with Wei Dai, Alessandro Miano and Freek Ruesink. This research was sponsored by the Army Research Office (ARO) under grant numbers W911NF-18-1-0212, W911NF-16-1-0349, and W911NF-23-1-0051,  and by the U.S. Department of  Energy, Office of Science, National Quantum Information Science Research Centers, Co-design Center for Quantum Advantage (C2QA) under contract number DE-SC0012704. The views and conclusions contained in this document are those of the authors and should not be interpreted as representing the official policies, either expressed or implied, of the U.S. Government. The U.S. Government is authorized to reproduce and distribute reprints for Government purposes notwithstanding any copyright notation herein.

%

\end{document}


    \title{Supplemental Material for "Fully Directional Quantum-limited Phase-Preserving Amplifier"}
    \author{G. Liu}
    \email{gangqiang.liu@yale.edu}
    \affiliation{Department of Applied Physics, Yale University, New Haven, CT 06520, USA}
    
    \author{A. Lingenfelter}
    \affiliation{Department of Applied Physics, Yale University, New Haven, CT 06520, USA}
    \affiliation{Department of Physics, University of Chicago, Chicago, IL 60637, USA}
    
    \author{V. R. Joshi}    
    \author{N. E. Frattini}
    \author{V. V. Sivak}
    \author{S. Shankar}
    \author{M. H. Devoret}
    \email{michel.devoret@yale.edu}
    \affiliation{Department of Applied Physics, Yale University, New Haven, CT 06520, USA}
    \date{\today}
    
\maketitle

\section{Minimal scattering matrix of 4-Port Fully Directional Amplifier (4PFDA)}
In this section, we will first derive the scattering matrix of a 4-port system that can perform fully directional phase-preserving amplification. Then, we will find the minimal form of such scattering matrices. Let's begin by summarizing the requirements on the scattering matrix for a fully directional phase-preserving amplifier.

For a linear amplifier with $N$ ports, the input and output signals are related by
\begin{align}
    \label{eq:S_general}
    \begin{bmatrix}
    a^{\rm{out}}_1 \\
    a^{\rm{out \dagger}}_1 \\
    \vdots \\
    a^{\rm{out}}_N \\
    a^{\rm{out \dagger}}_N \\    
    \end{bmatrix} = \tilde{S} 
    \begin{bmatrix}
    a^{\rm{in}}_1 \\
    a^{\rm{in \dagger}}_1 \\
    \vdots \\
    a^{\rm{in}}_N \\
    a^{\rm{in \dagger}}_N \\    
    \end{bmatrix}
\end{align}
where $a^{\rm{in}}_n$ ($a^{\rm{in} \dagger}_n$) and $a^{\rm{out}}_n$ ($a^{\rm{out} \dagger}_n$) are field operators of the input and output signals on the $n$-th port of the amplifier, and $\tilde{S}$ is the $2N$-by-$2N$ scattering matrix. For a fully directional phase-preserving amplifier, $\tilde{S}$ must satisfy the following conditions:
\begin{enumerate}[label=(\roman*)]
    \item {The scattering matrix can always be block-diagonalized on a proper mode basis,
    \begin{align}
    \label{eq:S_pp}
    L\tilde{S}L^{-1} =  
    \begin{bmatrix}
    S & 0 \\
    0 & S^*
    \end{bmatrix}
    \end{align}}
    where $L$ is the linear transformation from the mode basis in Eq.~(\ref{eq:S_general}) to the proper one. $S$ is a $N$-by-$N$ matrix which is often referred to as the scattering matrix even though it might mix $a_n$ and $a^\dagger_m$.   
    \item The scattering matrix is symplectic,
    \begin{align}
    \label{eq:S_symp}
        \tilde{S}^{T} \cdot J \cdot \tilde{S} = J
    \end{align}
    where $J$ is the symplectic matrix
    \begin{align}
    J=
    \begin{bmatrix}
        0 & 1 & & & \\
        -1 & 0 & & & \\
        & & \ddots & & \\
        & & & 0 & 1 \\
        & & & -1 & 0
    \end{bmatrix}.
    \end{align}
   This ensures that the device preserves the commutation relation of the input field operators \cite{Caves1982}.
   
    \item The sub-scattering matrix between the input port (Port 1) and output port (Port 2) should take the form
    \begin{align}
    \label{eq:S_directional}
        [S]_{1,2} = \begin{bmatrix}
        0 & 0 \\
        \sqrt{G}e^{i\phi} & 0
        \end{bmatrix}
    \end{align}
    where $S_{21} = \sqrt{G}e^{i\phi}$ is the gain from the input to the output port. This stipulates that both the input and output ports are matched, and the reverse isolation is infinity.
    \item Scattering matrix elements that couple other ports to the input and output ports should satisfy the following condition, 
    \begin{align}
    \label{eq:S_QL}
        \sum_{p=3}^{N} |S_{1p}|^2 = 1, \quad \sum_{p=3}^{N} |S_{2p}|^2 \approx G \ \rm{for} \ G\gg 1, 
    \end{align}
    such that both the back-action and added noise of the amplifier are quantum-limited when only vacuum noise enters the output port and all other auxiliary ports.
\end{enumerate} 

\subsection{Scattering matrix of 4PFDA}
In the case of a 4-port system, assuming Port 1 is the input port and Port 2 is the output port, we have the following possible mode bases that correspond to reduced scattering matrices of size $4$ by $4$
\begin{equation}
 \label{eq:mode_basis}
    A = 
    \begin{bmatrix}
    a_{1} \\
    a_{2} \\
    a_{3} \\
    a_{4}
    \end{bmatrix}, \quad
    \begin{bmatrix}
    a_{1} \\
    a_{2} \\
    a_{3} \\
    a^\dagger_{4}
    \end{bmatrix}, \quad
    \begin{bmatrix}
    a_{1} \\
    a^\dagger_{2} \\
    a_{3} \\
    a_{4}
    \end{bmatrix}, \quad    
    \begin{bmatrix}
    a_{1} \\
    a_{2} \\
    a^\dagger_{3} \\
    a^\dagger_{4}
    \end{bmatrix}, \quad    
    \begin{bmatrix}
    a_{1} \\
    a^\dagger_{2} \\
    a_{3} \\
    a^\dagger_{4}
    \end{bmatrix}, \quad \rm{or}
    \begin{bmatrix}
    a_{1} \\
    a^\dagger_{2} \\
    a^\dagger_{3} \\
    a^\dagger_{4}
    \end{bmatrix}.
\end{equation}
Each mode basis corresponds to a particular way that the four modes are coupled to each other.

The general form of the reduced scattering matrix $S$ for a 4-port system that perform fully directional amplification between Port 1 and Port 2 is 
\begin{equation}
    \label{eq:S_general}
    S = 
    \begin{bmatrix}
    0 & 0 & S_{13} & S_{14} \\
    S_{21} & 0 & S_{23} & S_{24} \\
    S_{31} & S_{32} & S_{33} & S_{34} \\
    S_{41} & S_{42} & S_{43} & S_{44}
    \end{bmatrix}
\end{equation}
with $|S_{21}|^2 > 1$ for amplification. Now our task is to determine, for each mode basis in Eq.~(\ref{eq:mode_basis}), whether there exists a scattering matrix of the form given by Eq.~(\ref{eq:S_general}) that satisfies Eq.~(\ref{eq:S_symp}), with
\begin{align}
    \label{eq:S_pp}
    \tilde{S} =  
    L^{-1} \begin{bmatrix}
    S & 0 \\
    0 & S^*
    \end{bmatrix} L 
\end{align}    
where $L$ is the linear transformation that transforms the general model basis $\tilde{A}$ 
\begin{align}
    \label{eq:A_general}
    \tilde{A} = \begin{bmatrix}
    a_1 \\
    a^{\dagger}_1 \\
    \vdots \\
    a_4 \\
    a^{\dagger}_4 \\    
    \end{bmatrix}     
\end{align}
into the mode basis on which $\tilde{S}$ is block diagonalized, namely,
\begin{align}
    \label{eq:L}
    \begin{bmatrix}
    A \\
    A^{\dagger} \\    
    \end{bmatrix} = L\tilde{A}     
\end{align}
with $A$ given in Eq.~(\ref{eq:mode_basis}). 

For mode basis 
\begin{equation}
    A = 
    \begin{bmatrix}
    a_{1} \\
    a_{2} \\
    a_{3} \\
    a_{4}
    \end{bmatrix}
\end{equation}
no scattering matrix exists that would satisfy all the conditions listed above. But we find there exists a set of scattering matrices with $|S_{21}|^2=1$, which represent 4-port circulators. Since we focus on the amplifier case in this work, we will not go into detail about the circulators here.

For the mode basis 
\begin{equation}
    A = 
    \begin{bmatrix}
    a_{1} \\
    a_{2} \\
    a_{3} \\
    a^\dagger_{4}
    \end{bmatrix}
\end{equation}
we get the following equations for the scattering matrix elements from Eq.~(\ref{eq:S_symp}),
\begin{eqnarray*}
|S_{32}|^2 - |S_{42}|^2 = 1 \\
|S_{21}|^2 + |S_{31}|^2 - |S_{41}|^2 = 1 \\
|S_{13}|^2 + |S_{23}|^2 + |S_{33}|^2 - |S_{43}|^2 = 1 \\
|S_{14}|^2 + |S_{24}|^2 + |S_{34}|^2 - |S_{44}|^2 = -1 \\
S_{31}S^*_{32} - S_{41}S^*_{42} = 0 \\
S_{32}S^*_{33} - S_{42}S^*_{43} = 0 \\
S_{32}S^*_{34} - S_{42}S^*_{44} = 0 \\
S_{21}S^*_{23} + S_{31}S^*_{33} - S_{41}S^*_{43} = 0 \\
S_{21}S^*_{24} + S_{31}S^*_{34} - S_{41}S^*_{44} = 0 \\
S_{13}S^*_{14} + S_{23}S^*_{24} + S_{33}S^*_{34} - S_{43}S^*_{44} = 0.
\end{eqnarray*}
These are the 10 equations for the 13 unknown scattering matrix elements. Therefore, three of these scattering matrix elements are independent variables. To solve these equations, we choose the following ansatzes 
\begin{eqnarray*}
    S_{21} &=& \sqrt{G_1}e^{i\theta_1}, \qquad S_{42} = \sqrt{G_2}e^{i\theta_2} \\
    S_{33} &=& \sqrt{\alpha_1}e^{i\phi_1}, \qquad S_{44} = \sqrt{\alpha_2}e^{i\phi_2}   
\end{eqnarray*}
where $\alpha_1, \alpha_2, G_2 \geq 0$, $G_1 \geq 1$. As we will show in the following, only 3 of these four ansatzes are independent of each other. Solving the set of equations above with these ansatzes, we obtain the other scattering matrix elements as
\begin{eqnarray*}
S_{13} &=& \sqrt{\alpha_1/G_1G_2+1}e^{i\theta_5}, \qquad S_{14} = \sqrt{\alpha_2/G_1(G_2+1)-1}e^{i\theta_6} \\
S_{23} &=& \sqrt{\alpha_1(G_1-1)/G_1G_2}e^{i(\theta_1+\theta_2-\theta_3-\theta_4+\phi_1)} \\
S_{24} &=& \sqrt{\alpha_2(G_1-1)/G_1(G_2+1)}e^{i(\theta_1-\theta_4+\phi_2)} \\
S_{31} &=& \sqrt{G_2(G_1-1)}e^{i(\theta_3-\theta_2+\theta_4)}, \qquad
S_{32} = \sqrt{G_2+1}e^{i\theta_3} \\
S_{34} &=& \sqrt{\alpha_2 G_2/(G_2+1)}e^{i(\theta_3-\theta_2+\phi_2)} \\
S_{41} &=& \sqrt{(G_1-1)(G_2+1)}e^{i\theta_4}, \qquad S_{43} = \sqrt{\alpha_1(G_2+1)/G_2}e^{i(\theta_2-\theta_3+\phi_1)}
\end{eqnarray*}
with the constrain giving by the last equation of the set as
\begin{equation}
    \sqrt{(\alpha_1 + G_1G_2)(\alpha_2-G_1(G_2+1))}e^{i(\theta_5-\theta_6)}=\sqrt{\alpha_1 \alpha_2}e^{i(\theta_2-\theta_3+\phi_1-\phi_2)}.
    \label{eq:A-Tamp-constrain}
\end{equation}
Without losing generality, we choose all the phases to be 0, namely, $\theta_n = 0$ for $n=1,2,...,6$ and $\phi_m = 0$ for $m=1,2$. Then we have 
\begin{equation}
    \sqrt{(\alpha_1 + G_1G_2)(\alpha_2-G_1(G_2+1))}=\sqrt{\alpha_1 \alpha_2}
    \label{eq:A-Tamp-constrain_1}
\end{equation}
and
\begin{equation}
    S = {\begin{bmatrix}
    0 & 0 & \sqrt{\alpha_1/G_1G_2 + 1} & \sqrt{\alpha_2/G_1(G_2+1)-1} \\
    \sqrt{G_1}  & 0 & \sqrt{\alpha_1(G_1-1)/G_1G_2} & \sqrt{\alpha_2(G_1-1)/G_1(G_2+1)} \\
    \sqrt{(G_1-1)G_2} &  \sqrt{G_2+1} & \sqrt{\alpha_1} & \sqrt{\alpha_2 G_2/(G_2+1)}  \\
     \sqrt{(G_1-1)(G_2+1)} & \sqrt{G_2} & \sqrt{\alpha_1(G_2+1)/G_2}  & \sqrt{\alpha_2} \\
    \end{bmatrix}}.
    \label{eq:S_T-amp}
\end{equation}


For the mode basis 
\begin{equation}
    A = 
    \begin{bmatrix}
    a_{1} \\
    a^\dagger_{2} \\
    a_{3} \\
    a^\dagger_{4}
    \end{bmatrix}, 
\end{equation}
we get the following equations for the scattering matrix elements from Eq.~(\ref{eq:S_symp}),
\begin{eqnarray*}
|S_{32}|^2 - |S_{42}|^2 = -1 \\
|S_{21}|^2 - |S_{31}|^2 + |S_{41}|^2 = -1 \\
|S_{13}|^2 - |S_{23}|^2 + |S_{33}|^2 - |S_{43}|^2 = 1 \\
|S_{14}|^2 - |S_{24}|^2 + |S_{34}|^2 - |S_{44}|^2 = -1 \\
S_{31}S^*_{32} - S_{41}S^*_{42} = 0 \\
S_{32}S^*_{33} - S_{42}S^*_{43} = 0 \\
S_{32}S^*_{34} - S_{42}S^*_{44} = 0 \\
S_{21}S^*_{23} - S_{31}S^*_{33} + S_{41}S^*_{43} = 0 \\
S_{21}S^*_{24} - S_{31}S^*_{34} + S_{41}S^*_{44} = 0 \\
S_{13}S^*_{14} - S_{23}S^*_{24} + S_{33}S^*_{34} - S_{43}S^*_{44} = 0.
\end{eqnarray*}
To solve these equations, we choose the following ansatzes
\begin{eqnarray*}
    S_{21} &=& \sqrt{G_1}e^{i\theta_1}, \qquad S_{42} = \sqrt{G_2+1}e^{i\theta_2} \\
    S_{33} &=& \sqrt{\alpha_1}e^{i\phi_1}, \qquad S_{44} = \sqrt{\alpha_2}e^{i\phi_2}.   
\end{eqnarray*}
With these ansatzes, we get
\begin{eqnarray*}
S_{13} &=& \sqrt{\alpha_1/G_1(G_2+1)+1}e^{i\theta_5}, \qquad S_{14} = \sqrt{\alpha_2/G_1G_2-1}e^{i\theta_6} \\
S_{23} &=& \sqrt{\alpha_1(G_1-1)/G_1G_2}e^{i(\theta_1+\theta_2-\theta_3-\theta_4+\phi_1)} \\ 
S_{24} &=& \sqrt{\alpha_2(G_1+1)/G_1G_2}e^{i(\theta_1-\theta_4+\phi_2)} \\
S_{31} &=& \sqrt{(G_1+1)(G_2+1)}e^{i(\theta_3-\theta_2+\theta_4)}, \qquad
S_{32} = \sqrt{G_2}e^{i\theta_3}, \\
S_{34} &=& \sqrt{\alpha_2 (G_2+1)/G_2}e^{i(\theta_3-\theta_2+\phi_2)} \\
S_{41} &=& \sqrt{G_2(G_1+1)}e^{i\theta_4}, \qquad S_{43} = \sqrt{\alpha_1G_2/(G_2+1)}e^{i(\theta_2-\theta_3+\phi_1)}
\end{eqnarray*}
with the constrain
\begin{equation}
    \sqrt{(\alpha_1 + G_1(G_2+1))(\alpha_2-G_1G_2)}e^{i(\theta_5-\theta_6)}=\sqrt{\alpha_1 \alpha_2}e^{i(\theta_2-\theta_3+\phi_1-\phi_2)}.
    \label{eq:A-Camp-constrain}
\end{equation}
Again, without losing generality, we can set all the phases to 0. Then we have 
\begin{equation}
    \sqrt{(\alpha_1 + G_1(G_2+1))(\alpha_2-G_1G_2)}=\sqrt{\alpha_1 \alpha_2},
    \label{eq:A-Camp-constrain}
\end{equation}
and
\begin{equation}
    S = {\begin{bmatrix}
    0 & 0 & \sqrt{\alpha_1/G_1(G_2+1) + 1} & \sqrt{\alpha_2/G_1G_2-1} \\
    \sqrt{G_1} & 0 & \sqrt{\alpha_1(G_1+1)/G_1(G_2+1)} & \sqrt{\alpha_2(G_1+1)/G_1G_2} \\
    \sqrt{(G_1+1)(G_2+1)} & \sqrt{G_2} & \sqrt{\alpha_1} & \sqrt{\alpha_2 (G_2+1)/G_2}  \\
     \sqrt{G_2(G_1+1)} & \sqrt{G_2+1} & \sqrt{\alpha_1G_2/(G_2+1)}  & \sqrt{\alpha_2} \\
    \end{bmatrix}}.
    \label{eq:S_C-amp}
\end{equation}

For other mode basis, no scattering matrix of the form given in Eq.~(\ref{eq:S_general}) can satisfy the requirement given Eq.~(\ref{eq:S_symp}). For example, for mode basis
\begin{equation}
    A = 
    \begin{bmatrix}
    a_{1} \\
    a^\dagger_{2} \\
    a_{3} \\
    a_{4}
    \end{bmatrix}
\end{equation}
the set of equations for the scattering matrix elements coming from Eq.~(\ref{eq:S_symp}) is
\begin{eqnarray*}
|S_{32}|^2 + |S_{42}|^2 = -1 \\
|S_{21}|^2 - |S_{31}|^2 - |S_{41}|^2 = -1 \\
|S_{13}|^2 - |S_{23}|^2 + |S_{33}|^2 + |S_{43}|^2 = 1 \\
|S_{14}|^2 - |S_{24}|^2 + |S_{34}|^2 + |S_{44}|^2 = 1 \\
S_{31}S^*_{32} + S_{41}S^*_{42} = 0 \\
S_{32}S^*_{33} + S_{42}S^*_{43} = 0 \\
S_{32}S^*_{34} + S_{42}S^*_{44} = 0 \\
S_{21}S^*_{23} - S_{31}S^*_{33} + S_{41}S^*_{43} = 0 \\
S_{21}S^*_{24} - S_{31}S^*_{34} - S_{41}S^*_{44} = 0 \\
S_{13}S^*_{14} - S_{23}S^*_{24} + S_{33}S^*_{34} + S_{43}S^*_{44} = 0
\end{eqnarray*}
of which 
\begin{equation}
    |S_{32}|^2 + |S_{42}|^2 = -1
\end{equation}
has no solution, since $|S_{nm}|^2 \ge 0$. This is the same situation for mode bases
\begin{equation}
    A = 
    \begin{bmatrix}
    a_{1} \\
    a_{2} \\
    a^\dagger_{3} \\
    a^\dagger_{4}
    \end{bmatrix},
    \begin{bmatrix}
    a_{1} \\
    a^\dagger_{2} \\
    a^\dagger_{3} \\
    a^\dagger_{4}
    \end{bmatrix}.
\end{equation}
Therefore, for a 4-port system, there exists two scattering matrices, as shown in Eq.~(\ref{eq:S_T-amp}) and (\ref{eq:S_C-amp}), which can perform fully directional phase-preserving amplification. The corresponding mode  base are
\begin{equation}
    A = 
    \begin{bmatrix}
    a_{1} \\
    a_{2} \\
    a_{3} \\
    a^\dagger_{4}
    \end{bmatrix},
    \begin{bmatrix}
    a_{1} \\
    a^\dagger_{2} \\
    a_{3} \\
    a^\dagger_{4}
    \end{bmatrix},
\end{equation}
respectively. We name them as the Trans- and Cis- 4-port fully directional amplifier (4PFDA), and will refer them as T-amp and C-amp for brevity. 

Following the same reasoning, one can show that for a system with fewer ports, no scattering matrix can satisfy all the requirements for fully directional phase-preserving amplification. Therefore, we conclude that {\it{the minimal construction of a fully directional phase-preserving amplifier requires a system of 4 ports.}} 

\subsection{Minimal scattering matrices of 4PFDAs}

For the scattering matrices we have found, which are given by Eq.~(\ref{eq:S_T-amp}) and Eq.~(\ref{eq:S_C-amp}), we now use the requirement of minimum back action and quantum-limited added noise, namely Eq.~(\ref{eq:S_QL}) to determine their minimal forms.

Assuming only vacuum noise enters the output port and the two ancillary ports, the back action of this amplifier, characterized by the number of noise photons sent back to the input source, is  
\begin{align}
\label{eq:T-4PFDA_ba}
    N^{\rm{ba}}_T = \frac{1}{2} \left(\frac{\alpha_1}{G_1G_2} + \frac{\alpha_2}{G_1(G_2+1)} \right)
\end{align}
and the added noise when referring back to its input port is,
\begin{align}
\label{eq:T-4PFDA_add}
    N^{\rm{add}}_T = \frac{1}{2} \frac{G_1-1}{G_1} \left(\frac{\alpha_1 }{G_1 G_2} + \frac{\alpha_2}{G_1(G_2+1)} \right).
\end{align}
From Eq.~(\ref{eq:A-Tamp-constrain_1}) we have 
\begin{align}
    \alpha_1 \ge 0, \quad \alpha_2 \ge G_1(G_2+1),
\end{align}
thus the back action and added noise of the amplifier reach their minimum values when $\alpha_1 = 0$, $\alpha_2 = G_1(G_2+1)$:
\begin{align}
\label{eq:T-4PFDA_ba_min}
    N^{\rm{ba}}_T = \frac{1}{2}
\end{align}
\begin{align}
    N^{\rm{add}}_T = \frac{1}{2} \frac{G_1-1}{G_1}.
\end{align}
In the large gain limit, $G_1\gg1$, the added noise reaches the quantum limit as:
\begin{align}
    N^{\rm{add}}_T = \frac{1}{2} \left(1 - \mathcal{O}(\frac{1}{G_1}) \right).
\end{align}
Therefore, the scattering matrix of T-amp with minimum back action and quantum-limited added noise is
\begin{widetext}
\begin{align}
\label{eq:S_T-4PFDA_QL}
    S_{T} = {\begin{bmatrix}
    0 & 0 & 1 & 0 \\
    \sqrt{G_1} & 0 & 0 & \sqrt{G_1-1} \\
    \sqrt{(G_1-1)G_2} & \sqrt{G_2+1} & 0 & \sqrt{G_1G_2} \\
    \sqrt{(G_1-1)(G_2+1)} & \sqrt{G_2} & 0 & \sqrt{G_1(G_2+1)} \\
    \end{bmatrix}}
\end{align}
\end{widetext}
which can be further simplified, by setting $G_2=0$, to the minimum form of
\begin{align}
\label{eq:S_T-4PFDA_QL_min}
    S_{T} = {\begin{bmatrix}
    0 & 0 & 1 & 0 \\
    \sqrt{G_1} & 0 & 0 & \sqrt{G_1-1} \\
    \sqrt{G_1-1} & 1 & 0 & 0 \\
    0 & 1 & 0 & \sqrt{G_1} \\
    \end{bmatrix}}.
\end{align}
This result shows that in its simplest form, the T-amp is an amplifier with matched input and output ports and forward gain, and perfect reverse isolation between these two ports. One of the two auxiliary ports is matched while the other has reflection gain.

Similarly, for C-amp the minimum form of its scattering matrix with quantum-limited back action and added noise, obtained from Eq.~(\ref{eq:S_C-amp}) under the condition that 
\begin{align}
    \alpha_1 = 0, \quad \alpha_2 = (G_1+1)(G_2+1), \quad G_2 = 0,
\end{align}
is 
\begin{align}
\label{eq:S_C-4PFDA_QL_min}
    S_{C} = {\begin{bmatrix}
    0 & 0 & 1 & 0 \\
    \sqrt{G_1} & 0 & 0 & \sqrt{G_1+1} \\    
    \sqrt{G_1+1} & 0 & 0 & \sqrt{G_1} \\
    0 & 1 & 0 & 0 \\
    \end{bmatrix}}.
\end{align}
The corresponding back action and added noise, in the large gain limit $G_1\gg 1$, are
\begin{align}
\label{eq:C-4PFDA_ba_min}
    N^{\rm{ba}}_C = \frac{1}{2}
\end{align}
\begin{align}
\label{eq:C-4PFDA_add_min}
    N^{\rm{add}}_C = \frac{1}{2} \left(1 + \mathcal{O}(\frac{1}{G_1+1})\right).
\end{align}
This scattering matrix shows that in its simplest form, the C-amp is an amplifier with all its ports matched.

\subsection{From scattering matrix to mode couplings of 4PDAs}
In this section, we will construct the coupling Hamiltonian of the 4-port full directional phase-preserving amplifiers from their scattering matrices.

\subsubsection{Matrix form of Hamiltonian}
In order to introduce the matrix form of the Hamiltonian, we need to write the coupling terms in a symmetric form. For a system of $N$-coupled modes, the Hamiltonian is
\begin{equation}
    \frac{H}{\hbar}=\sum^N_{m=1}\left(\omega_m a^\dagger_m a_m + \frac{1}{2}\sum_{n \in C_m} (g_{mn}a^\dagger_m a_n + g^*_{mn}a_m a^\dagger_n) + \frac{1}{2}\sum_{p \in G_m} (g_{mp}a^\dagger_m a^\dagger_p + g^*_{mp}a_m a_p) \right)
\end{equation}
where $C_m$ ($G_m$) represents the set of modes that coupled to $a_m$ through photon conversion (gain) process with couplings strength $g_{mn}$ ($g_{mp}$). In the case of parametrically coupled systems, $g_{mn} = \tilde{g}_{mn}e^{-i(\omega_m - \omega_n)t}$, $g_{mp} = \tilde{g}_{mp}e^{-i(\omega_m + \omega_p)t}$, with $\tilde{g}_{mn}$ and $\tilde{g}_{mp}$ are complex coupling strength. Also $g_{mn} = g^*_{nm}$ and $g_{mp} = g_{pm}$, since they are the coupling strengths for the same pair of modes, respectively.

The equation of motion (EOM) of the field operator is given by the Langevin equation
\begin{equation}
    \frac{d a_m}{dt} = \frac{i}{\hbar}[H,a_m] - \frac{\kappa_m}{2}a_m + \sqrt{\kappa_m}a^{\rm{in}}_m(t)
\end{equation}
and the input and output fields are related by the input-output relation
\begin{equation}
    \sqrt{\kappa_m}a_m = a_m^{\rm{in}}(t) + a_m^{\rm{out}}(t).
\end{equation}
From the Hamiltonian, we have
\begin{equation}
    \frac{i}{\hbar}[H, a_m] = -i\omega_m a_m - \frac{i}{2}\sum_m g_{mn}a_m - \frac{i}{2}\sum_n g_{ln}a^\dagger_m
\end{equation}
Substituting these relations into the EOM, we get
\begin{align}
      & (\frac{\kappa_m}{2} + \frac{d}{dt} + i\omega_m )a^{\rm{out}}_m + \frac{i}{2}\sum_{n \in C_m}\sqrt{\frac{\kappa_m}{\kappa_n}}g_{mn}a^{\rm{out}}_n + \frac{i}{2}\sum_{p \in G_m}\sqrt{\frac{\kappa_m}{\kappa_p}}g_{mp}a^{\rm{out} \dagger}_p \\ 
      & = (\frac{\kappa_m}{2} - \frac{d}{dt} - i\omega_m)a^{\rm{in}}_m - \frac{i}{2}\sum_{n \in C_m}\sqrt{\frac{\kappa_m}{\kappa_n}}g_{mn}a^{\rm{in}}_n - \frac{i}{2}\sum_{p \in G_m}\sqrt{\frac{\kappa_m}{\kappa_p}}g_{mp}a^{\rm{in} \dagger}_p
\end{align}
Assuming the input signals are all monochromatic, then the input and output field operators can be written as,
\begin{equation}
    a^{\rm{in, out}}_m = a^{\rm{in, out}}_m[\omega^s_m] e^{-i\omega^s_m t}, \qquad a^{\rm{in, out} \dagger}_m = a^{\rm{in, out}}_m[\omega^s_m]^\dagger e^{i\omega^s_m t}
\end{equation}
and
\begin{align}
      & \left(\frac{\kappa_m}{2} - i(\omega^s_m - \omega_m)\right)a^{\rm{out}}_m[\omega^s_m] + \frac{i}{2}\sum_{n \in C_m}\sqrt{\frac{\kappa_m}{\kappa_n}}\tilde{g}_{mn}a^{\rm{out}}_n[\omega^s_n] + \frac{i}{2}\sum_{p \in G_p}\sqrt{\frac{\kappa_m}{\kappa_p}}\tilde{g}_{mp}a^{\rm{out} \dagger}_p[\omega^s_p] \\ 
      & = \left(\frac{\kappa_m}{2} + i(\omega^s_m - \omega_m) \right)a^{\rm{in}}_m[\omega^s_m] - \frac{i}{2}\sum_{n \in C_m}\sqrt{\frac{\kappa_m}{\kappa_n}}\tilde{g}_{mn}a^{\rm{in}}_n[\omega^s_n] - \frac{i}{2}\sum_{p \in G_m}\sqrt{\frac{\kappa_m}{\kappa_p}}\tilde{g}_{mp}a^{\rm{in} \dagger}_p[\omega^s_p]
\end{align}
which can also be obtained by Fourier transforming of the time-domain EOM. Now we can introduce the matrix form of the EOM of the system with
\begin{equation}
    \Sigma = 
    \begin{bmatrix}
    \kappa_1/2 & & & \\
    & \cdot & & \\
    &  & \cdot & \\
    & & & \kappa_N/2
    \end{bmatrix}
\end{equation}
\begin{equation}
    M =
    \begin{bmatrix}
    i\Delta_1 & M_{12} & \cdots & M_{1N} \\
    M_{21} & i\Delta_2 & \cdots & M_{2N} \\
    \vdots &  \vdots & \cdots & \vdots \\
    M_{N1} & M_{N2} & \cdots & i\Delta_N
    \end{bmatrix}
\end{equation}
where  
\begin{equation}
\Delta_m = \left\{ \begin{array}{ll}
    \omega^s_m - \omega_m & \textrm{for $a_m$}\\
    -(\omega^s_m - \omega_m) & \textrm{for $a^\dagger_m$}
  \end{array} \right.
\end{equation}
and 
\begin{equation}
\label{eq:M_g}
M_{mn} = \left\{ \begin{array}{ll}
    -\frac{i}{2} \sqrt{\frac{\kappa_m}{\kappa_n}}(\tilde{g}_{mn} + \tilde{g}^*_{nm}) & \textrm{for $n \in C_m$}\\
    \\
    -\frac{i}{2} \sqrt{\frac{\kappa_m}{\kappa_n}} (\tilde{g}_{mn} + \tilde{g}_{nm}) & \textrm{for $n \in G_m$}
  \end{array} \right.
\end{equation}
Because $\tilde{g}_{mn} = \tilde{g}^*_{nm}$ for coupling between $a_m$ and $a_n$, $\tilde{g}_{mn} = \tilde{g}_{nm}$ for coupling between $a_m$ and $a^\dagger_n$,
\begin{equation}
    M_{mn} = -i \sqrt{\frac{\kappa_m}{\kappa_n}}\tilde{g}_{mn}
\end{equation}
and
\begin{equation}
M_{nm} = \left\{ \begin{array}{ll}
    - \frac{\kappa_n}{\kappa_m} M^*_{mn} & \textrm{for $n \in C_m$}\\
    \\
    \frac{\kappa_n}{\kappa_m} M^*_{mn} & \textrm{for $n \in G_m$}
  \end{array} \right.
\end{equation}
The relation between input and output field now can be written as
\begin{equation}
    A^{\rm{out}} = (\Sigma+M)^{-1}(\Sigma-M)A^{\rm{in}}
\end{equation}
from which we can define the scattering matrix as
\begin{equation}
    \label{eq:Z_to_S}
    S = (\Sigma+M)^{-1}(\Sigma-M)
\end{equation}
and the full generalized scattering matrix of the system is 
\begin{equation}
    \tilde{S} = \begin{bmatrix}
    S & \\
    & S^*
    \end{bmatrix}
\end{equation}

From Eq.~\ref{eq:Z_to_S}, we have
\begin{equation}
    \label{eq:S_to_Z}
    M = \Sigma(I-S)(I+S)^{-1}
\end{equation}
which gives the coupling between models for a given scattering matrix.

\subsubsection{Mode coupling of 4PFDAs}


From Eq.~(\ref{eq:S_to_Z}) and the minimal scattering matrices of the T-amp (Eq.~(\ref{eq:S_T-4PFDA_QL_min})) and C-amp (Eq.~(\ref{eq:S_C-4PFDA_QL_min})), we get the model coupling matrix for these two amplifiers as
\begin{equation}
M_T = \frac{1}{2}
\begin{bmatrix}
0 & \kappa_1 & -\kappa_1 & -\frac{\sqrt{G-1}}{\sqrt{G}+1} \kappa_1\\
-\kappa_2 & 0 & \kappa_2 &  -\frac{\sqrt{G-1}}{\sqrt{G}+1} \kappa_2\\
\kappa_3 & -\kappa_3 & 0 & \frac{\sqrt{G-1}}{\sqrt{G}+1} \kappa_3 \\
-\frac{\sqrt{G-1}}{\sqrt{G}+1} \kappa_4 & -\frac{\sqrt{G-1}}{\sqrt{G}+1} \kappa_4 & \frac{\sqrt{G-1}}{\sqrt{G}+1} \kappa_4 & 0
\end{bmatrix},
\label{eq:A-M_T}
\end{equation} and
\begin{equation}
M_C = \frac{1}{2}
\begin{bmatrix}
0 & \frac{\sqrt{G}+1}{\sqrt{G-1}} \kappa_1 & -\kappa_1 & -\frac{\sqrt{G}+1}{\sqrt{G-1}} \kappa_1\\
\frac{\sqrt{G}+1}{\sqrt{G-1}} \kappa_2 & 0 & -\frac{\sqrt{G}+1}{\sqrt{G-1}} \kappa_2 &  \kappa_2\\
\kappa_3 & -\frac{\sqrt{G}+1}{\sqrt{G-1}} \kappa_3 & 0 & \frac{\sqrt{G}+1}{\sqrt{G-1}} \kappa_3 \\
-\frac{\sqrt{G}+1}{\sqrt{G-1}} \kappa_4 & -\kappa_4 & \frac{\sqrt{G}+1}{\sqrt{G-1}} \kappa_4 & 0
\end{bmatrix}
\label{eq:A-M_C}
\end{equation}
where $G=G_1$ for T-amp and $G=G_1+1$ for C-amp. This choice of definition ensures that $G$ will correspond to the reflection gain of a single photon gain coupling as we will show shortly.

Notice that the diagonal elements of the mode coupling matrices are all zeros, which means the scattering matrices we found are for the special case of resonant input signals. The behavior/performance of the amplifiers for arbitrary input frequency can be obtained by adding the detuning terms back to the diagonal elements of the coupling matrix from which the scattering matrix can then be calculated. In the next section, we will study the frequency dependent behavior of the amplifiers.

These coupling matrices show that there are 6 couplings in both amplifiers, even though some ports are isolated in the scattering matrix. The amplitude and phase of these coupling, calculated from Eq.~(\ref{eq:M_g}), are listed in the following table.
\begin{widetext}
\begin{equation}
\left.\begin{array}{c|c|c}
 & \rm{T-amp} & \rm{C-amp} \\\hline
(g_{12}, \phi_{12}) & \displaystyle{( \frac{\sqrt{\kappa_1 \kappa_2}}{2}, \frac{\pi}{2})} & \displaystyle{( \frac{\sqrt{\kappa_1 \kappa_2}}{2} \frac{\sqrt{G}+1}{\sqrt{G-1}}, -\frac{\pi}{2})} \\\hline 
(g_{13}, \phi_{13}) & \displaystyle{( \frac{\sqrt{\kappa_1 \kappa_3}}{2}, -\frac{\pi}{2})} & \displaystyle{( \frac{\sqrt{\kappa_1 \kappa_3}}{2}, \frac{\pi}{2})} \\\hline 
(g_{14}, \phi_{14}) & \displaystyle{( \frac{\sqrt{\kappa_1 \kappa_4}}{2} \frac{\sqrt{G-1}}{\sqrt{G}+1}, -\frac{\pi}{2})} & \displaystyle{( \frac{\sqrt{\kappa_1 \kappa_4}}{2} \frac{\sqrt{G}+1}{\sqrt{G-1}}, -\frac{\pi}{2} )} \\\hline 
(g_{23}, \phi_{23}) & \displaystyle{( \frac{\sqrt{\kappa_2 \kappa_3}}{2}, \frac{\pi}{2})} & \displaystyle{( \frac{\sqrt{\kappa_2 \kappa_3}}{2} \frac{\sqrt{G}+1}{\sqrt{G-1}}, -\frac{\pi}{2} )} \\\hline 
(g_{24}, \phi_{24}) & \displaystyle{( \frac{\sqrt{\kappa_2 \kappa_4}}{2} \frac{\sqrt{G-1}}{\sqrt{G}+1}, -\frac{\pi}{2})} & \displaystyle{( \frac{\sqrt{\kappa_2 \kappa_4}}{2}, \frac{\pi}{2})} \\\hline 
(g_{34}, \phi_{34}) & \displaystyle{( \frac{\sqrt{\kappa_3 \kappa_4}}{2} \frac{\sqrt{G-1}}{\sqrt{G}+1}, \frac{\pi}{2})} & \displaystyle{( \frac{\sqrt{\kappa_3 \kappa_4}}{2} \frac{\sqrt{G}+1}{\sqrt{G-1}}, \frac{\pi}{2})}
\end{array}\right.
\label{eq:coupling table}
\end{equation}
\end{widetext}

With the mode coupling matrix and the mode basis, we can now reconstruct the full Hamiltonian of these amplifiers. For T-amp, we get 
\begin{align*}
    \frac{H_T}{\hbar} & = \sum_{n=1}^{4}\omega_n a^\dagger_n a_n + g_{12}\left(a^\dagger_1 a_2 e^{-i(\Omega_{12}t+\phi_{12})} + h.c.\right) + g_{13}\left(a^\dagger_1 a_3 e^{-i(\Omega_{13}t+\phi_{13})} + h.c.\right) \\
    & \quad + g_{14}\left(a^\dagger_1 a^\dagger_4 e^{-i(\Omega_{14}t+\phi_{14})} + h.c. \right) + g_{23}\left(a^\dagger_2 a_3 e^{-i(\Omega_{23}t+\phi_{23})} + h.c. \right) \\
    & \quad + g_{24}\left(a^\dagger_2 a^\dagger_4 e^{-i(\Omega_{24}t+\phi_{24})} + h.c. \right) + g_{34}\left(a^\dagger_3 a^\dagger_4 e^{-i(\Omega_{34}t+\phi_{34})} + h.c. \right) 
\end{align*}
where $\Omega_{mn}$ and $\phi_{mn}$ are the frequency and phase of the parametric drive that couples mode $a_k$ and $a_l$. The frequencies of these parametric drives are
\begin{equation}
    \Omega_{12} = \omega_1 - \omega_2, \quad \Omega_{13} = \omega_1 - \omega_3, \quad \Omega_{23} = \omega_2 - \omega_3, \\ \quad
    \Omega_{14} = \omega_1 + \omega_4, \quad \Omega_{24} = \omega_2 + \omega_4, \quad \Omega_{34} = \omega_3 + \omega_4.
\end{equation}
For C-amp, we get
\begin{align*}
    \frac{H_C}{\hbar} & = \sum_{n=1}^{4}\omega_n a^\dagger_n a_n + g_{12}\left(a^\dagger_1 a^\dagger_2 e^{-i(\Omega_{12}t+\phi_{12})} + h.c.\right) + g_{13}\left(a^\dagger_1 a_3 e^{-i(\Omega_{13}t+\phi_{13})} + h.c.\right) \\
    & \quad + g_{14}\left(a^\dagger_1 a^\dagger_4 e^{-i(\Omega_{14}t+\phi_{14})} + h.c. \right) + g_{23}\left(a^\dagger_2 a^\dagger_3 e^{-i(\Omega_{23}t+\phi_{23})} + h.c. \right) \\
    & \quad + g_{24}\left(a^\dagger_2 a_4 e^{-i(\Omega_{24}t+\phi_{24})} + h.c. \right) + g_{34}\left(a^\dagger_3 a^\dagger_4 e^{-i(\Omega_{34}t+\phi_{34})} + h.c. \right) 
\end{align*}
where the parametric drive frequencies are
\begin{equation}
    \Omega_{12} = \omega_1 + \omega_2, \quad \Omega_{13} = \omega_1 - \omega_3, \quad \Omega_{23} = \omega_2 + \omega_3, \\ \quad
    \Omega_{14} = \omega_1 + \omega_4, \quad \Omega_{24} = \omega_2 - \omega_4, \quad \Omega_{34} = \omega_3 + \omega_4.
\end{equation}

From the Hamiltonian of the amplifiers, we can identify the couplings as either photon gain coupling or photon conversion coupling. For T-amp, there are three photon gain couplings -- $g_{14}$, $g_{24}$, $g_{34}$ -- with photon number gain of $G$ in reflection, and three photon conversion couplings -- $g_{12}$, $g_{13}$, $g_{23}$ -- with perfect conversion ($C=1$) between coupled modes. For C-amp, there are four photon gain couplings -- ($g_{12}$, $g_{14}$, $g_{23}$ and $g_{34}$) -- with photon gain of $G$ in reflection, and two photon conversion couplings -- $g_{12}$, $g_{24}$ -- with perfect conversion between coupled modes. These coupling schemes are graphically represented in Fig.3 of the main text.

It is important to point out that the direct coupling between the two auxiliary modes is crucial for achieving quantum-limited added noise. Without this coupling, the noise photons will be amplified by a larger gain than signal photons leading to sever degradation in SNR. For such a system -- 4 modes with 5 couplings -- to achieve quantum-limited noise performance, it has to be operated in the situation whether there is unity reflection on both the input and output port \cite{Metelmann2017}.

\section{4PFDA performance vs frequency}
In this section, we will study the performance of T-amp and C-amp for input signal at arbitrary frequency. We will first study the ideal case of perfect conversions between modes. Then, we will show amplifier performance with imperfect conversions and show that these amplifiers provides better performance than existing multi-parametric direction amplifiers under practical operation conditions.

To obtain the frequency dependent scattering matrix for the 4PDAs with perfect conversions, we simply need to add the detuning terms back to the diagonal elements of the coupling matrices shown in Eq.~(\ref{eq:A-M_T}) and Eq.~(\ref{eq:A-M_C}). For an input signal at frequency $\omega_s = \omega_1 + \Delta$ to the input port, the mode coupling matrix for T-amp is
\begin{equation}
M_T[\Delta] = \frac{1}{2}
\begin{bmatrix}
-i\Delta & \kappa_1 & -\kappa_1 & -\frac{\sqrt{G-1}}{\sqrt{G}+1} \kappa_1\\
-\kappa_2 & -i\Delta & \kappa_2 &  -\frac{\sqrt{G-1}}{\sqrt{G}+1} \kappa_2\\
\kappa_3 & -\kappa_3 & -i\Delta & \frac{\sqrt{G-1}}{\sqrt{G}+1} \kappa_3 \\
-\frac{\sqrt{G-1}}{\sqrt{G}+1} \kappa_4 & -\frac{\sqrt{G-1}}{\sqrt{G}+1} \kappa_4 & \frac{\sqrt{G-1}}{\sqrt{G}+1} \kappa_4 & -i\Delta
\end{bmatrix},
\label{eq:A-M_T-freq}
\end{equation}
where $\Delta_1=\Delta_2=\Delta_3=\Delta, \Delta_4 = -\Delta$, and for C-amp is
\begin{equation}
M_C[\Delta] = \frac{1}{2}
\begin{bmatrix}
-i\Delta & \frac{\sqrt{G}+1}{\sqrt{G-1}} \kappa_1 & -\kappa_1 & -\frac{\sqrt{G}+1}{\sqrt{G-1}} \kappa_1\\
\frac{\sqrt{G}+1}{\sqrt{G-1}} \kappa_2 & -i\Delta & -\frac{\sqrt{G}+1}{\sqrt{G-1}} \kappa_2 &  \kappa_2\\
\kappa_3 & -\frac{\sqrt{G}+1}{\sqrt{G-1}} \kappa_3 & -i\Delta & \frac{\sqrt{G}+1}{\sqrt{G-1}} \kappa_3 \\
-\frac{\sqrt{G}+1}{\sqrt{G-1}} \kappa_4 & -\kappa_4 & \frac{\sqrt{G}+1}{\sqrt{G-1}} \kappa_4 & -i\Delta
\end{bmatrix}
\label{eq:A-M_C-freq}
\end{equation}
where $\Delta_1=\Delta_3=\Delta, \Delta_2=\Delta_4=-\Delta$. Then the scattering matrix can be calculated from these results together with
\begin{equation}
\Sigma =
\begin{bmatrix}
\kappa_1/2 & & & \\
& \kappa_2/2 & & \\
&  & \kappa_3/2 & \\
& & & \kappa_4/2
\end{bmatrix}
\end{equation}
according to Eq.~(\ref{eq:Z_to_S}). The scattering matrix elements are now complex functions of all the parameters ($G$, $\kappa_n$'s and $\Delta$). In order to simplify the discussion, let's consider the special case of all the modes having the same linewidth, namely $\kappa_1=\kappa_2=\kappa_3=\kappa_4=\kappa$. In this case, the scattering matrix elements between the input port (Port 1) and output port (Port 2) are
\begin{eqnarray}
S^{T}_{11}[\delta], S^{T}_{22}[\delta] = \frac{i\delta+3\delta^2-2i(1+\sqrt{G})\delta^3-2(1+\sqrt{G})\delta^4}{1-i(4+\sqrt{G})\delta-3(2+\sqrt{G})\delta^2+4i(1+\sqrt{G})\delta^3+2(1+\sqrt{G})\delta^4} \\
S^{T}_{12}[\delta] = \frac{\delta(i+\delta+\sqrt{G}\delta)}{1-i(4+\sqrt{G})\delta-3(2+\sqrt{G})\delta^2+4i(1+\sqrt{G})\delta^3+2(1+\sqrt{G})\delta^4} \\
S^{T}_{21}[\delta] = -\frac{(i+\delta)(i\sqrt{G}+\delta+\sqrt{G}\delta)}{1-i(4+\sqrt{G})\delta-3(2+\sqrt{G})\delta^2+4i(1+\sqrt{G})\delta^3+2(1+\sqrt{G})\delta^4}
\end{eqnarray} 
for T-amp, and 
\begin{eqnarray}
S^{C}_{11}[\delta], S^{C}_{22}[\delta] = \frac{\delta(i(\sqrt{G}+1)+(\sqrt{G}+3)\delta+2i(\sqrt{G}-1)\delta^2+2(\sqrt{G}-1)\delta^3)}{1-i4\delta-2(\sqrt{G}-3)\delta^2-4i(\sqrt{G}-1)\delta^3-2(\sqrt{G}-1)\delta^4} \\
S^{C}_{12}[\delta] = \frac{\sqrt{G-1}\delta^2}{1-i4\delta-2(\sqrt{G}-3)\delta^2-4i(\sqrt{G}-1)\delta^3-2(\sqrt{G}-1)\delta^4} \\
S^{C}_{21}[\delta] = \frac{\sqrt{G-1}(i+\delta)^2}{1-i(4+\sqrt{G})\delta-3(2+\sqrt{G})\delta^2+4i(1+\sqrt{G})\delta^3+2(1+\sqrt{G})\delta^4}
\end{eqnarray}
for C-amp, where $\delta=\Delta/\kappa$ is the normalized detuning. These scattering matrix elements are plotted as solid curves in Fig.~4 of the main text. Similarly, we can obtain the scattering matrix elements for the case of $C<1$, which are plotted as dashed curves in Fig.~4 of the main text.  

\begin{figure}[htb]
    \centering
    \includegraphics[width=0.8\linewidth]{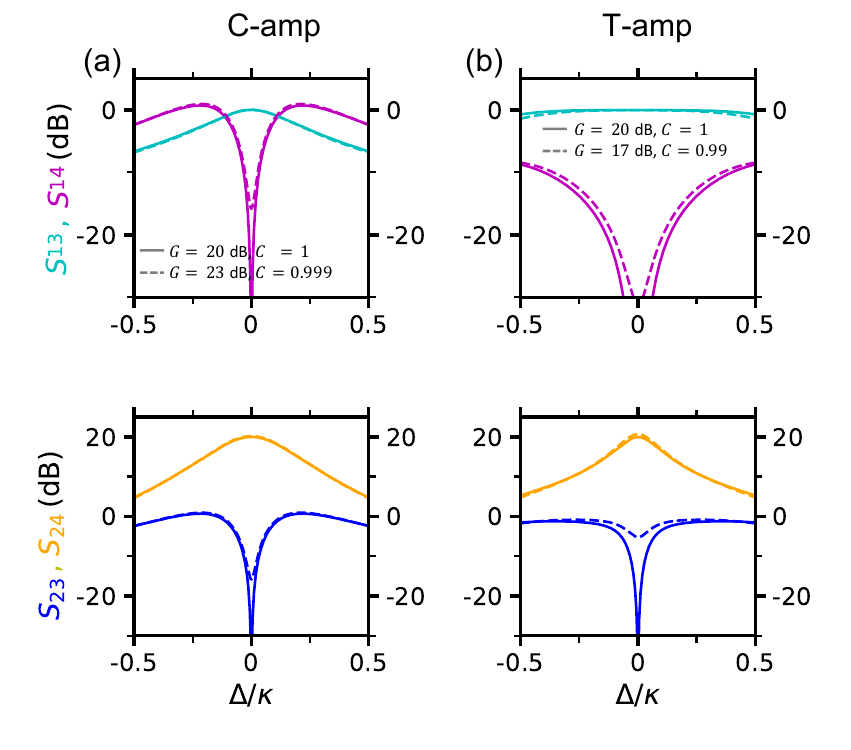}
    \caption{Back action and added noise of 4PFDAs. Back action ($S_{13}$, $S_{14}$) and added noise ($S_{23}$, $S_{24}$) of (a) C-amp for gain coupling of $G=20$~dB and perfect conversion $C=1$ (solid line), and $G=23$~dB and $C=0.999$ (dashed line); (b) T-amp for gain coupling of $G=20$~dB and perfect conversion $C=1$ (solid line), and $G=23$~dB and $C=0.999$ (dashed line).  }
    \label{fig:backaction_noise} 
\end{figure}

The scattering matrix elements between the auxiliary ports and the input and output ports are shown in Fig.~\ref{fig:backaction_noise}, which shows that both amplifiers retain quantum-limited noise performance and minimal back action for near resonance input signals.

\section{Relation between 4PFDA and 3-port directional amplifiers (3PDAs)}

\begin{figure}[htb]
    \centering
    \includegraphics[width=0.5\linewidth]{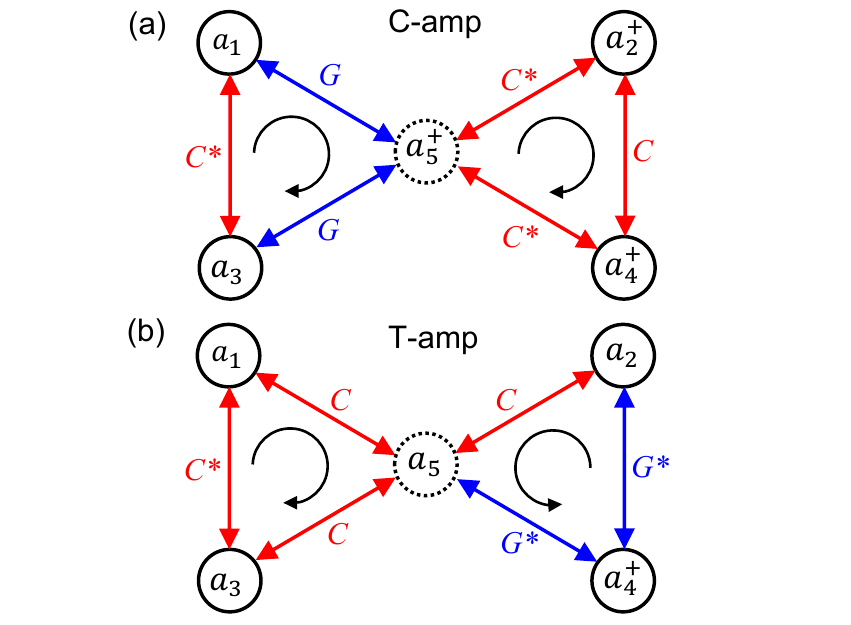}
    \caption{Equivalent coupling graph of 4PFDAs with a 3-port ciculator and a 3-port directional amplifier. (a) The equivalent coupling graph of the C-amp. It consists of a 3-mode directional amplifier on the input side couples to a 3-mode circulator through a shared mode ($a_5$) on the output side. (b) The equivalent coupling graph of the T-amp. It consists of a 3-mode circulator on the input side couples to a 3-mode directional amplifier through a shared mode ($a_5$) on the outupt side. The circular arrows indicate the directionality of each 3-mode loop.}
    \label{fig:4PFDA-3PDA} 
\end{figure}

The mode coupling graph of the T-amp and the C-amp (Fig.3 of main text) can be decomposed into two 3-mode coupling graphs that share a common node as shown in Fig.~\ref{fig:4PFDA-3PDA}, respectively. This illustrates the relation between the 4PFDAs and 3PDAs which are previously implemented in superconducting circuits \cite{Abdo2014,Sliwa2015,Lecocq2017} and opto-mechanical systems \cite{Fang2017}. The T-amp is equivalent to coupling a 3PDA and a 3-port circulator with the input signal entering from a port of the 3PDA and leaving from a port of the circulator. The C-amp represents the configuration in which the input signal enters the system from one port of the circulator and leaving from a port of the 3PDA.

\begin{figure}[htb]
    \centering
    \includegraphics[width=0.5\linewidth]{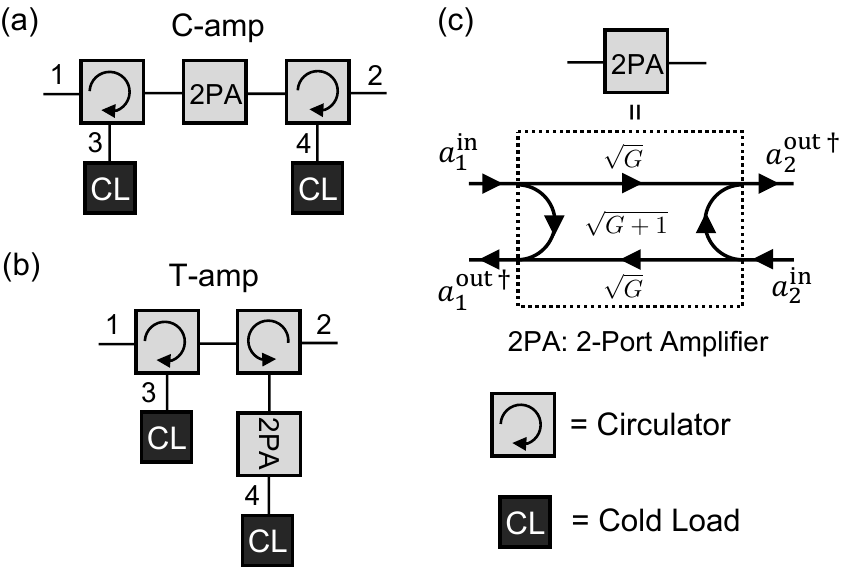}
    \caption{Equivalent construction of 4PDA with circulators and 2-port amplifier. (a) Two circulators with a 2-port amplifier (2PA) in between from the equivalent directional amplifier of a all-matched 4PFDA. (b) Equivalent construction with two circulators and a 2-port amplifier of acilla-unmatched 4PFDA. (c) Scattering graph of a 2-port amplifier with amplitude gain $\sqrt{G+1}$ in reflection and $\sqrt{G}$ in transmission powered by a single pump at frequency $\omega_p \sim \omega_1+\omega_2$. }
    \label{fig:4PFDA-circulator-2PA} 
\end{figure}

The T-amp and C-amp can also be constructed with a 2-port amplifier and two 3-port circulators as shown in Fig.~\ref{fig:4PFDA-circulator-2PA}. For T-amp, it corresponds to the situation where the 2-port amplifier is used in reflection, while for C-amp it is used in transmission. In fact, the configuration that equivalent to T-amp has been widely used in qubit readout with 2-port amplifiers.

